\documentclass[final,3p,times,onecolumn]{elsarticle}

\usepackage{mathrsfs}
\usepackage{amssymb,amsmath,amsthm}
\usepackage{amsfonts}
\usepackage{booktabs}
\usepackage{graphicx}
\usepackage{multicol}
\usepackage{slashbox}
\usepackage{color}
\usepackage{rotating}
\usepackage{algorithm}
\usepackage{algorithmic}
\usepackage{epstopdf}
\usepackage{array}

\journal{Applied Energy}

\begin{document}
	
	\begin{frontmatter}
		
		
		\title{Short-term load forecasting using optimized LSTM networks based on EMD}
		
		\author[label1]{Tiantian Li}
		\author[label1]{Bo Wang\corref{cor1}}
		\author[label1]{Min zhou}
		\author[label2]{Junzo Watada}
		
		\cortext[cor1]{Corresponding author.}
		
		\address[label1]{School of Management and Engineering, Nanjing University, Nanjing 210093, China}
		\address[label2]{Graduate School of Information, Production and
			Systems, Waseda University, Kitakyushu 808-0135, Japan}

		\begin{abstract}
			Short-term load forecasting is one of the crucial sections in smart grid. Precise forecasting enables system operators to make reliable unit commitment and power dispatching decisions. With the advent of big data, a number of artificial intelligence techniques such as back propagation, support vector machine have been used to predict the load of the next day. Nevertheless, due to the noise of raw data and the randomness of power load, forecasting errors of existing approaches are relatively large. In this study, a short-term load forecasting method is proposed on the basis of empirical mode decomposition and long short-term memory networks, the parameters of which are optimized by a particle swarm optimization algorithm. Essentially, empirical mode decomposition can decompose the original time series of historical data into relatively stationary components and long short-term memory network is able to emphasize as well as model the timing of data, the joint use of which is expected to effectively apply the characteristics of data itself, so as to improve the predictive accuracy. The effectiveness of this research is exemplified on a realistic data set, the experimental results of which show that the proposed method has higher forecasting accuracy and applicability, as compared with existing methods.

		\end{abstract}
		
		
	\end{frontmatter}
	
	\section{Introduction}
	
	Based on historical data, power load forecasting is to explore the developing law of electricity, establish models between power demand and features, then make a valid prediction of future load~\cite{Intro1}. 
	A lot of operations in power systems sharply depend on the future information provided by predictions, for example making a satisfying unit commitment (UC) decision~\cite{Mean}, saving energy and reducing the cost of power generation~\cite{Mean2}. Therefore, accurate load forecasting has far-reaching significance for this industry.
	
	In the past decades, researchers have made great efforts on load forecasting. Divided from the length of the predictive period, load forecasting includes long-term load forecasting (LTLF) usually a few months to years~\cite{Intro3}, medium-term load forecasting (MTLF) usually a few weeks to months~\cite{Intro4} and short-term load forecasting (STLF) usually one hour to several weeks~\cite{Intro5}. In reality, due to the in-depth study and extensive application of the day-ahead scheduling by system operators, STLF has a relatively larger application market. In the literature, there have been various approaches developed for STLF, which can be divided into three categories: 1) The traditional statistical methods represented by autoregressive integrated moving average model (ARIMA)~\cite{ARIMA1}, Kalman filter method~\cite{KF} and autoregressive based time varying model~\cite{AR1}. 2) The artificial intelligence (AI) methods such as extreme learning machine (ELM)~\cite{ELM1}, generalized regression neural network (GRNN)~\cite{GRNN1} and support vector regression (SVR)~\cite{SVR1}. 3) The hybrid methods such as hybridizing extended Kalman Filter (EKF) and ELM ~\cite{HB1}.
	
	For the statistical methods, Lee and Ko utilized lifting schemes and ARIMA models, in which the original load series was decomposed into separate sub-series at different revolution levels to enhance the frequency characteristic. The proposed method was tested on practical load data from Taipower Company~\cite{ARIMA1}. 
	Guan et al. presented a hybrid Kalman filters-based method, in which extended Kalman filters and unscented Kalman filters were utilized for low frequency and high frequency components respectively. The method was exemplified on a data set from ISO New England, which demonstrates the effectiveness of the proposed method in capturing different features of load components~\cite{KF}.
	Vu et al. proposed an autoregressive based time varying model to forecast electricity demand in a short-term period, coefficients of which were updated at pre-set time intervals. The model was tested on the data for the state of New South Wales, Australia, and the performance of the autoregressive based time varying model is proved to outperform conventional seasonal autoregressive and neural network models~\cite{AR1}.
	
	Although the traditional methods mentioned above are relatively simple and easy to implement, their prediction confidence is limited. With the increasing demand of forecasting accuracy and the rapid development of AI algorithms, recent studies have been focused on the AI-based approaches.
	Li et al. proposed a novel ensemble method for STLF based on wavelet transform, ELM and partial least squares regression. Experimental results on actual load data from ISO New England show the effectiveness of the proposed method~\cite{ELM1}.
	Xiao et al. presented a modified GRNN based on a multi-objective firefly algorithm (MOFA), the validity of which is justified by experiments based on the half-hourly electrical load data from three states in Australia~\cite{GRNN1}.
	Chen et al. adopted SVR for the prediction of daily power, in which load for four typical office buildings  was utilized as experimental sample data. It is proved that the SVR model offers a higher degree of prediction accuracy and stability in STLF compared to the other traditional forecasting models~\cite{SVR1}.
	
	Generally, the above AI-based methods have significantly improved the forecasting accuracy. However, there still exists difficulties in setting parameters, speeding up training-rate and avoiding local minima. Therefore, researchers have considered to combine traditional methods with AI algorithms.
	Liu et al. utilized a model which hybrids EKF and ELM to predict the hourly power demand. The proposed method is proved to be effective through tests on typical micro-grids~\cite{HB1}. 
	
	Objectively, the above hybrid method mitigates the instability of learning characteristics of a single method. However, few of existing methods including the above hybrid one consider the inherent nature of load data, which is a long-term series with undulation multiplicity and periodic variety. Fortunately, recurrent neural network (RNN) could be an impactful tool to deal with such sequences due to its connections among hidden nodes~\cite{Intro6}. Initially, RNN was put forward by Schmidhuber in 1992. Recently, various studies have been focused on the application of RNN. 
	Cheng et al. proposed a global RNN-based sentiment method for sentiment classifications of Chinese micro blogging texts~\cite{RNN2}.
	Cho et al. presented a novel neural network model called RNN Encoder-Decoder which consists of two RNN to improve the performance of a statistical machine translation system~\cite{RNN3}.
	Shi et al. proposed a novel pooling-based deep recurrent neural network (PDRNN) for household load forecasting. The experiments on 920 smart metered customers from Ireland confirm the outperformance of PDRNN~\cite{RNN41}.
	However, the main disadvantage of applying RNN to load forecasting is that the activation function of RNN uses chain rules to operate the gradient descent algorithm. Generally, there may exist many continuous multiplications for the items less than one, which will lead to the problem of gradient vanishing.
	To mitigate the defect, long short-term memory (LSTM) network has been proposed by Hochreiter and Schmidhuber in 1997~\cite{LSTM1}. As an improved version of RNN, LSTM changes the internal structure and transfers the state of hidden layers through the concept of gates. Thus, LSTM can effectively alleviate the gradient vanishing of RNN. Recently, LSTM has been extensively used in various forecasting problems. 
	Zhao et al. presented a novel traffic forecast model based on LSTM networks, which considered the temporal-spatial correlation in traffic system through a two-dimensional network composed of many memory units~\cite{LSTM2}.
	Kong et al. proposed a framework based on LSTM to forecast the power load of a single energy user, the effectiveness of which was tested on a publicly available set of real residential data~\cite{LSTM3}.
	
	Although LSTM for load forecasting is not new, few of them take into account the pre-processing of raw data and the optimization of LSTM parameters. Thus, there still exists a relatively large promotion space of forecasting accuracy. 
	In this paper, a method that combines empirical mode decomposition (EMD), particle swarm optimization (PSO) and LSTM is proposed to forecast the next 24 hour load profile, named EMD-PSO-LSTM. In particular, this method only uses load data as the feature and label for training, which not only effectively reduces the dimension of required data, but also excavates the information of load data. 
	From the literature, EMD is an adaptive approach of time-frequency analysis proposed by Huang in 1998~\cite{EMD1}. The core idea of EMD is consistent with Fourier transform (FT)~\cite{FT} and wavelet transform (WT)~\cite{WT}, which is plausible to decompose the signal into a superposition of independent components. However, EMD decomposes the signal based on the time-scaled characteristic of data itself and requires no basis function, which is different from FT and WT. 
	Accordingly, in the decomposition of nonlinear and non-stationary signals, EMD has a more superior signal-noise ratio than FT and WT. 
	Based on the above advantages, this study first uses EMD to decompose the original sequence of load data into a finite number of relatively stable components, through which the periodicity of each component is highlighted. In view of the similar periodicity of certain components, the ingredients can be recombined into several frequency-parts through experimental experience. 
	Second, for each frequency-part, a separate LSTM network is constructed to predict sub-load values of the next 24 hours. Synchronously, PSO is utilized to optimize the input and output weights of LSTM to reduce the random error. Experimental results on a realistic data set demonstrate that the prediction obtained by the proposed method is confident to achieve higher accuracy than existing approaches. 
	
	The contribution of this study is two-fold: First, by decomposing the original load data into similar frequency components through EMD, LSTM can expediently learn the characteristic of the processed data. Secondly, the parameters of LSTM are optimized by PSO, which improves the forecasting accuracy evidently.
	
	The remainder of this paper is arranged as follows: In Section II, the knowledge of RNN, LSTM and EMD is elaborated. Section III gives a detailed description of EMD-PSO-LSTM. In Section IV, a number of experiments were performed to demonstrate the effectiveness of the proposed method, and detailed discussions on the experimental results are provided. Finally, Section V summarizes conclusions and expectations for future research.
	
	\section{Preliminaries}
	In this section, we briefly introduce the knowledge of RNN, LSTM and EMD.
	\subsection{Recurrent neural network}
	
	RNN is an artificial neural network specially designed to solve the problem of certain recursion structures in daily phenomena~\cite{RNN}. In RNN, there exists a logistic function $tanh$ as described in Equation (~\ref{t}), which maps all values between -1 and 1. The standard RNN operates the input sequence $I$ through $tanh$ to get the output state $h$ at each time step $t$,
	\begin{equation}
	tanh(x)= \frac{e^{x}-e^{-x}}{e^{x}+e^{-x}}\label{t}.
	\end{equation} 
	
	\begin{figure}[h]
		
		\centering
		\includegraphics[scale=0.15]{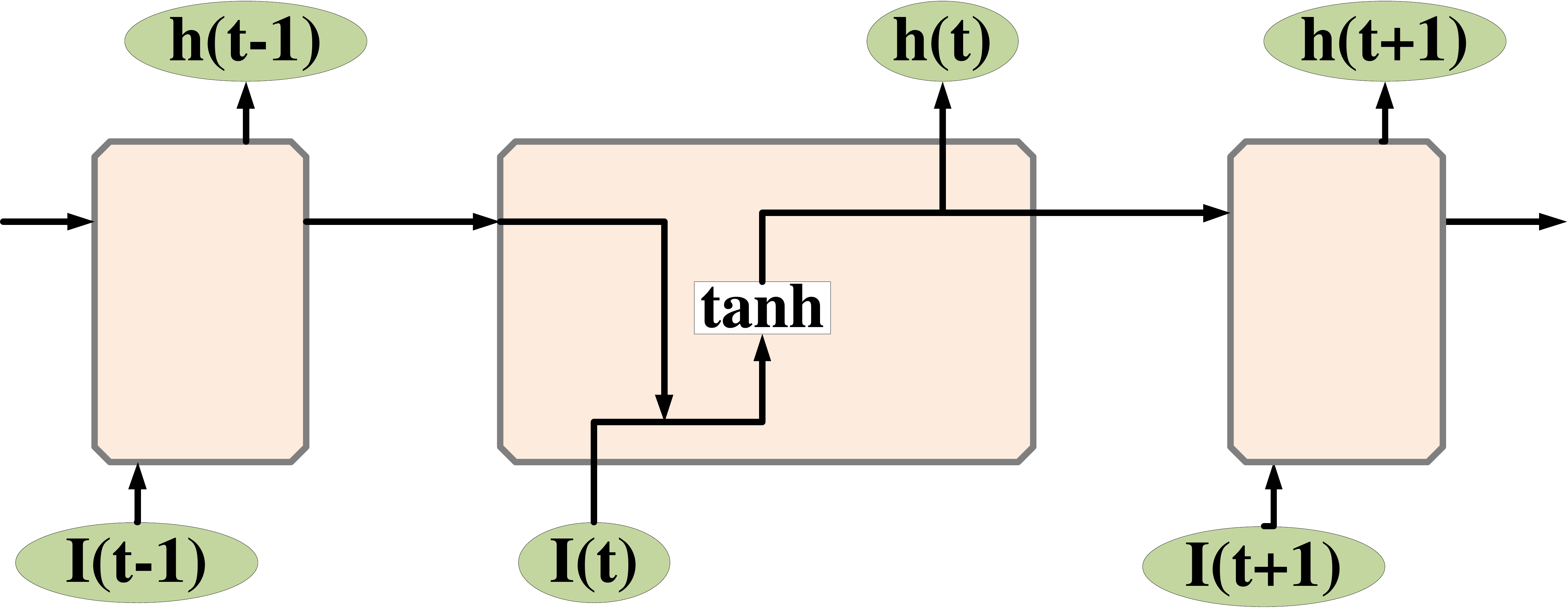}
		\caption{Structure of RNN}
		\label{fig:RNN1}
	\end{figure}
	
	Specifically, RNN not only learns the information $I_t$ of current time $t$, but also relies on the previous sequence information $h_{t-1}$. As shown in Fig. 1, RNN can be intuitively treated as multiplications of the same neural network. Similar to back propagation (BP), RNN utilizes the back propagation through time (BRTT) algorithm to reverse the propagation error as well, but the process of RNN is related to time. Owing to the special structure with chain features, RNN solves the problem of information preservations, and has unique advantages in dealing with time series.
	
	\subsection{Long short-term memory network}
	
	Although RNN is an impactful tool in dealing with recursion problems, there still exists problems such as gradient vanishing. Generally, when there has a relatively small value in the matrix of RNN, multiple matrix multiplications will decrease the gradient at an exponential rate and eventually disappear after a few steps. As the time interval increases, the network is unable to learn long-distant information so as to arise the problem of gradient vanishing.
	Fortunately, there are already solutions to the gradient vanishing problem, among which LSTM is the most common method~\cite{LSTM1}, as shown in Fig. 2. 
	\begin{figure}[h]
		
		\centering
		\includegraphics[scale=0.15]{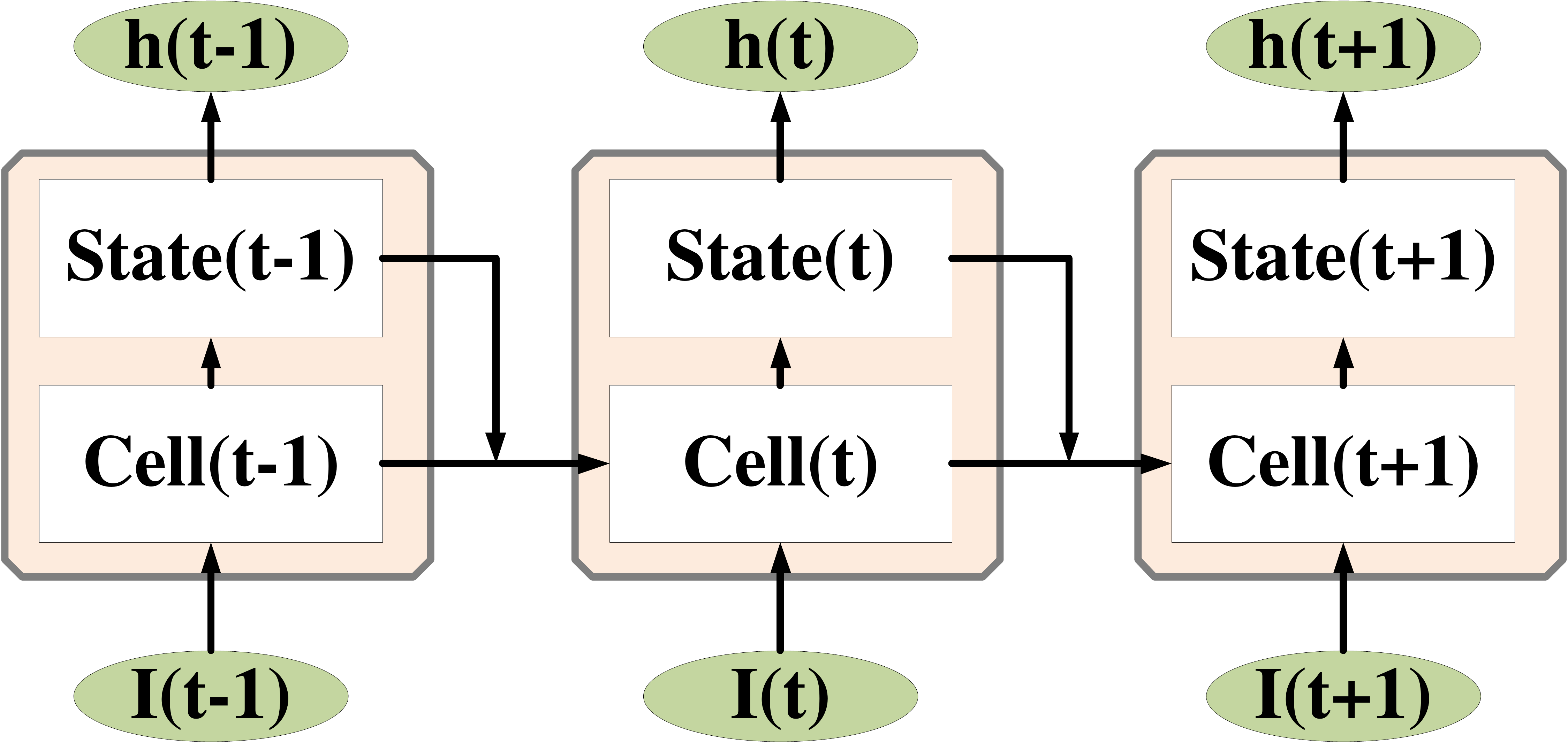}
		\caption{Unfolded LSTM}
		\label{fig:LSTM1}
	\end{figure}
	
	Generally, LSTM inherits the feature of RNN that each hidden layer interconnects with the others, but the connection of back-nodes to front-nodes of LSTM is improved by using three gates, i.e. forget gate, input gate and output gate. Then the input and output of each cell are selected by the gates, as shown in Fig. 3. 
	
	\begin{figure}[h]
		
		\centering
		\includegraphics[scale=0.135]{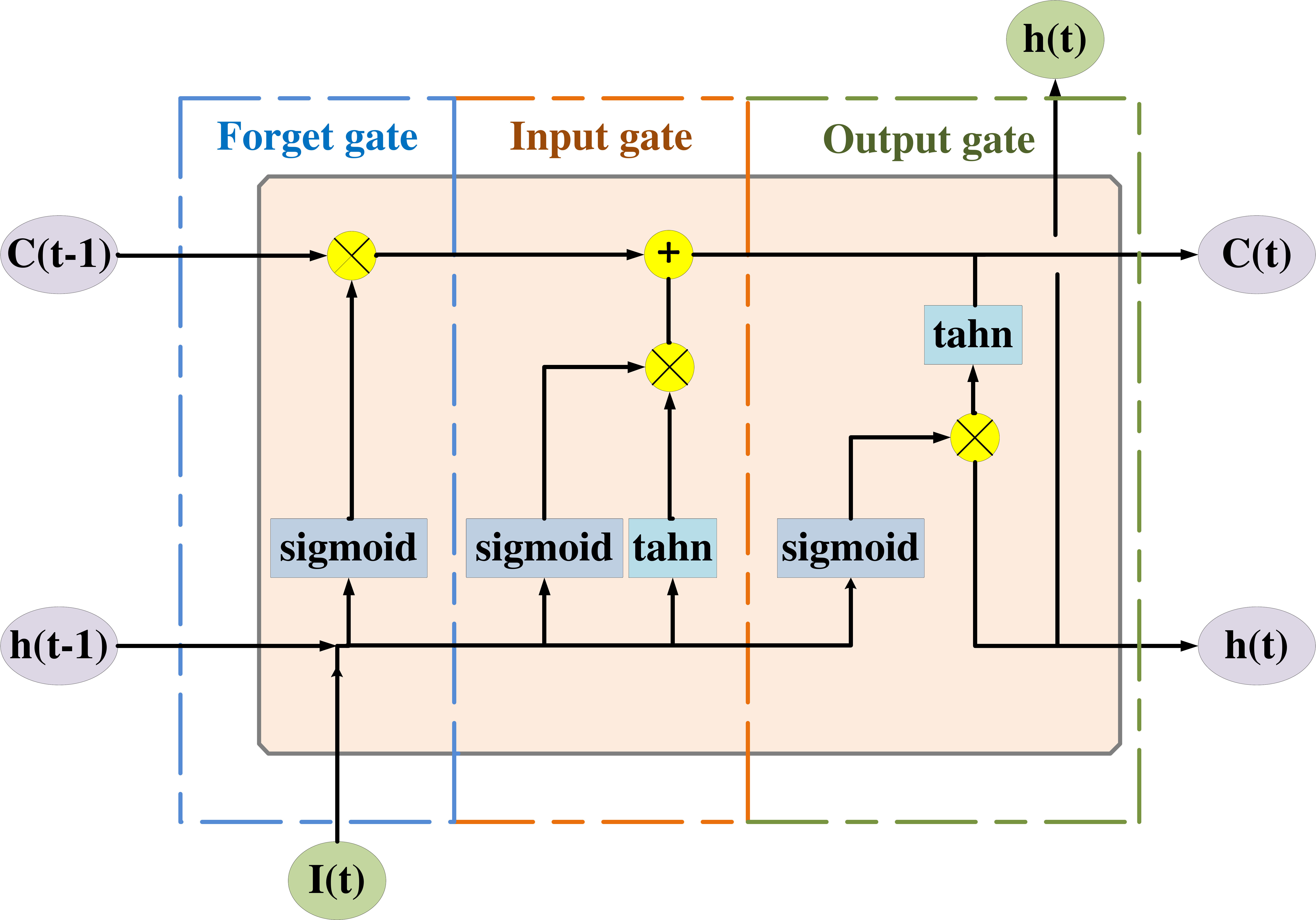}
		\caption{Unit of LSTM}
		\label{fig:LSTM2}
	\end{figure}
	
	Specifically, the forget gate determines the information to be discarded from a cell, which includes a logistic function $sigmoid$ as described in Equation (~\ref{s}) and a bitwise multiplication, so as to transfer information selectively,
	
	\begin{equation}
	sigmoid(x)=\frac{1}{1+e^{-x}}\label{s},
	\end{equation}
	\begin{equation}
	f_t=\sigma(W_{xf}\cdot x_t+W_{hf}\cdot h_{t-1}+bias_f)\label{1}.
	\end{equation}
	
	According to Equation (~\ref{1}), the forget gate accepts the current input $x_t$, the previous state $h_{t-1}$ and the correction $bias_f$ of the network. Then the gate calculates a temporary output between 0 and 1, which means the degree of forgetting with corresponding weights $W$ for each part in the unit state.
	
	Then, the input gate determines the information to be retained from a cell, which can be split into two portions,
	\begin{equation}
	i_t=\sigma(W_{xi}\cdot x_t+W_{hi}\cdot h_{t-1}+bias_i)\label{2},
	\end{equation}
	\begin{equation}
	c_t=\tau(W_{xc}\cdot x_t+W_{hc}\cdot h_{t-1}+bias_c)\label{3},
	\end{equation}
	where Equation (~\ref{2}) decides the value to be updated, and Equation (~\ref{3}) creates a tanh layer $\tau$ containing alternative values which can be added to a new state.
	
	After completing the above parts, the state of LSTM is updated as follows,
	\begin{equation}
	C_t=f_t*C_{t-1}+i_t*c_t\label{4}.
	\end{equation}
	Obviously, Equation (~\ref{4}) connects the pre-state $C_{t-1}$ and the present temporary-state $c_t$.
	
	Finally, based on the cell condition, LSTM outputs the selected state through the output gate,
	\begin{equation}
	o_t=\sigma(W_{xo}\cdot x_t+W_{ho}\cdot h_{t-1}+bias_o)\label{5},
	\end{equation}
	\begin{equation}
	h_t=o_t*(\tau(C_t))\label{6},
	\end{equation}
	where Equation (~\ref{5}) runs a sigmoid layer to determine the unit state section to be exported, and Equation (~\ref{6}) deals current output $o_t$ and state $C_t$ with a tanh layer to write a new hidden layer state.
	
	Beyond the above LSTM, there are also some other derivatives of RNN, such as peephole LSTM~\cite{PP}, gate recurrent unit (GRU)~\cite{GRU} and grid LSTM~\cite{GL}. Readers are encouraged to refer to the related studies for the detailed knowledge.
	
	\subsection{Empirical mode decomposition}
	
	Generally, the original data contains several intrinsic mode functions (IMFs). With IMFs and a residual waveform (Res) overlapping, the raw data is rebuilt as a composite one,
	\begin{equation}
	S_{org}=\sum{IMFs}+Res\label{7},
	\end{equation}
	where the decomposed IMF components contain local characteristics of different time scales of the original data, and for every IMF:
	\begin{enumerate}
		\item Over the whole time range, the difference between the number of local extremes and zero crossing points cannot exceed one.
		\item The average of local maximum's upper envelope and local minimum's lower envelope must be 0.
	\end{enumerate}
	
	To obtain the above IMFs, the raw data is decomposed with EMD, which is based on the following hypothesis:
	\begin{enumerate}
		\item The original data has at least two extreme points, i.e. a maximum value and a minimum value.
		\item The time scale of characteristics is defined by the time interval between two extreme points.
		\item If there exists no extreme points but deformation points, extreme points can be obtained by differentiating the data once or several times. Then the decomposition result can be obtained by the integration.
	\end{enumerate}
	
	Normally, the procedure of EMD is as below:
	\begin{enumerate}
		\item Initialize $r_0=x(t)$, $i=1$, where $x(t)$ is the original data series.
		\item Get $IMF_i$ 
		\begin{enumerate}
			\item Initialize $h_0=r_{i-1}(t),j=1$.
			\item Find out all extreme points of $h_{j-1}(t)$.
			\item Apply Cubic Spline Interpolation on maximum and minimum extreme values of $h_{j-1}(t)$ to establish the upper $e_{max}(t)$ and lower $e_{min}(t)$ envelopes.
			\item Calculate the average value as $m_{j-1}(t)=(e_{min}(t)+e_{min}(t))/2$.
			\item Calculate $h_j(t)=h_{j-1}(t)-m_{j-1}(t)$.
			\item If $h_j(t)$ is IMF, then $imf_i(t)=h_j(t)$. Else, $j=j+1$, turn to Step b).		
		\end{enumerate}
		\item Calculate $r_i(t)=r_{i-1}(t)-imf_i(t)$.
		\item If extreme points of $r_i(t)$ are more than two, then $i=i+1$, repeat Steps 2) and 3). Else, the decomposition is completed with $r_i(t)$ as Res.
	\end{enumerate}
	
	In summary, EMD first seeks the difference in each part of the mixed sequence, then separates IMFs with the same frequency. Eventually, the raw data could be rebuilt as Equation (~\ref{7}). As a mixed sequence, power load data can be divided into several frequency parts as well. The high frequency component describes the slight change of the load curve, and the low frequency component depicts the amplitude. Therefore, it is reasonable to use EMD to analyze the raw data of STLF problems.
	
	\section{Proposed method}
	As shown in Fig. 4, the proposed EMD-PSO-LSTM mainly consists four portions:
	
	\begin{enumerate}
		\item Pre-data processing with EMD and reorganization
		\item Modeling of LSTM network
		\item Parameter optimization with PSO 
		\item Renormalization and linear addition
	\end{enumerate}
	
	\begin{figure}[h]
		
		\centering
		\includegraphics[width=8.5cm,height=14cm]{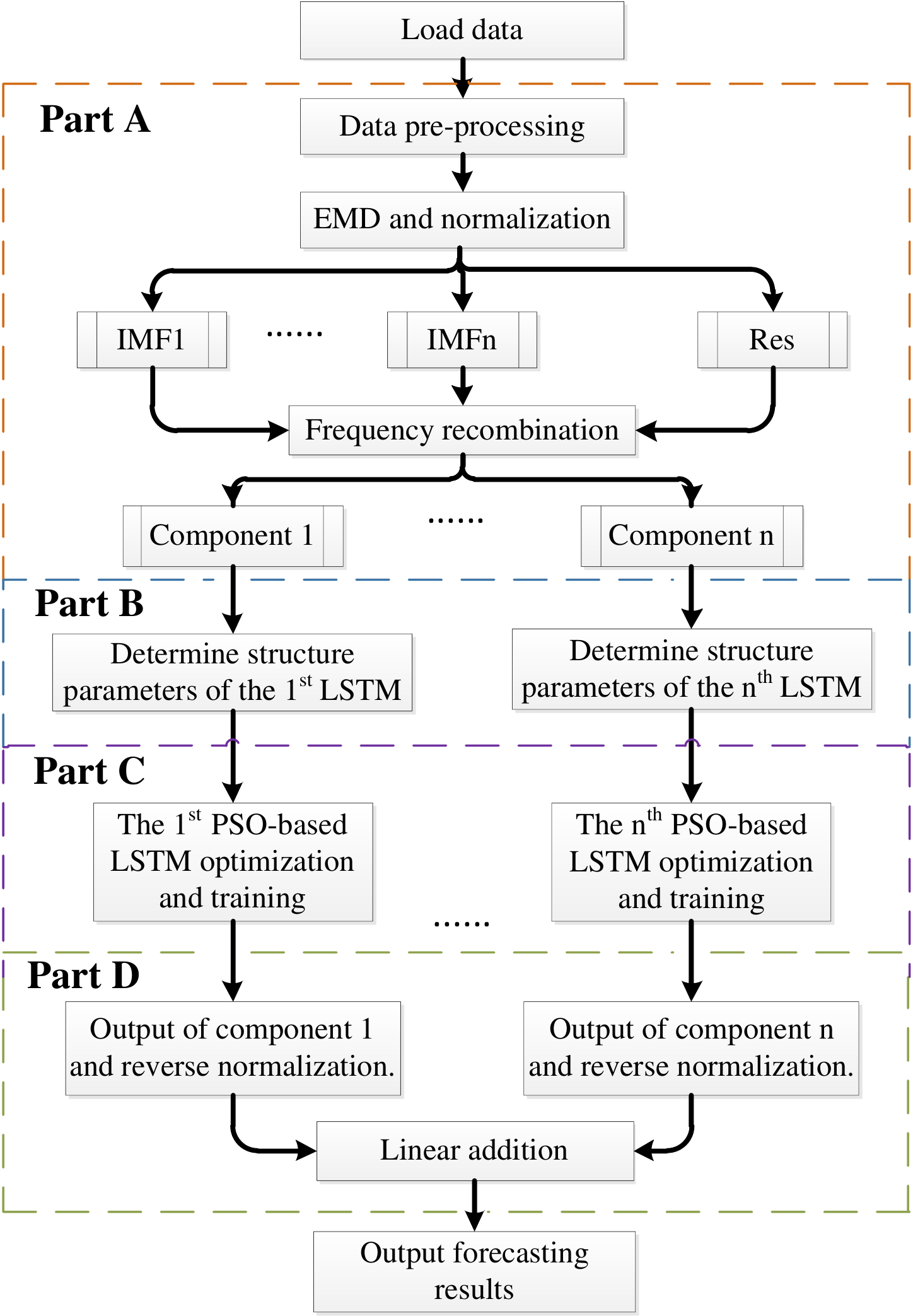}
		\caption{Flow chart of the proposed EMD-PSO-LSTM}
		\label{fig:LSTM3}
	\end{figure}
	
	\subsection{Pre-data processing with EMD and reorganization}
	In reality, the original data may contain bad data such as defaults and outliers due to equipment failures or operational errors, which could have a sharp effect on the forecasting accuracy. Thus, the input data could be pre-cleaned according to the following steps before training LSTM.
	
	\textbf{Step 1}: Calculate the average $\mu$ and the variance $\sigma$ of the raw load data.
	
	\textbf{Step 2}: Filter bad data based on the 3$\sigma$ principle with threshold $\varepsilon$ as Equation (~\ref{8}) and (~\ref{9}).
	\begin{equation}
	P(\mu-3\sigma,\mu+3\sigma)=0.9974\label{8},
	\end{equation}
	
	\begin{equation}
	\mid X-\mu \mid >3\sigma\varepsilon\label{9}.
	\end{equation}
	
	\textbf{Step 3}: Revise bad data as:
	\begin{equation}
	X_{a,b}= \frac{\alpha}{2}\sum{X_{a\pm{1},b}}+\frac{\beta}{2}\sum{X_{a,b\pm{1}}}+\gamma{\mu}\label{10},
	\end{equation}
	where in Equation (~\ref{10}), $\alpha+\beta+\gamma=1$, $X_{a,b}$ is the revised point, $X_{a\pm{1},b}$ and $X_{a,b\pm{1}}$ are the horizontal and vertical load points.
	
	After recognizing and processing of bad data, EMD could be performed to obtain several IMFs and one Res. Then the maximum $X_{max}$ and minimum $X_{min}$ are obtained for each component. Finally, the processed data is normalized based on Equation (~\ref{11}) and all load data is allocated between 0 and 1,
	\begin{equation}
	X_{a,b}=\frac{X_{a,b}-X_{min}}{X_{max}-X_{min}}\label{11}.
	\end{equation}
	
	After pretreatment, each component has its own unique frequency characteristic. Considering the interaction among components, certain components can be recombined to not only reduce the final predictive error but also lighten the computation burden. 
	
	\subsection{Modeling of LSTM network}
	
	Generally, this paper is to predict the future 24 hour load value. Proverbially, the load data shows periodic changes such as weekly and quarterly, which should be exploited so as to determine a suitable input pattern. 
	Appropriate input and label could enhance the validity of a input pattern. Taking the seven-to-one input-pattern as an example, if system operators intend to predict the 24 hour load value on March 8, they would train the model by using the historical load data from February 1 to February 7 as an input, with the past load on February 8 as an output label. Then the historical load from February 2 to February 8 would be used as the next input with the load of February 9 as the next label. In summary, the $N$-to-one training pattern is to iteratively predict the load values of the $(N+1)$ day by using the load values of the previous $N$ days as features, until the load forecast for the target day is obtained.
	\begin{algorithm}[htbp]  
		\caption{Adaptive moment estimation}  
		\begin{algorithmic}
			\renewcommand{\algorithmicrequire}{\textbf{Require:}}
			\REQUIRE  $\alpha$: Step size. 
			\renewcommand{\algorithmicrequire}{\textbf{Require:}}
			\REQUIRE  $\beta_1,\beta_2$ $\in$ [0,1): Exponential decay rates for the moment estimates
			\renewcommand{\algorithmicrequire}{\textbf{Require:}}
			\REQUIRE  $f(\theta)$: Stochastic objective function with parameter $\theta$
			\renewcommand{\algorithmicrequire}{\textbf{Require:}}
			\REQUIRE  $\theta_0$: Initialize parameter vector;
			\STATE  $m_0$ $\gets$ 0: Initialize the first moment vector;
			\STATE  $v_0$ $\gets$ 0: Initialize the second moment vector; 
			\STATE  $t$ $\gets$ 0: Initialize time step; 
			\WHILE  {$\theta_t$ not converged} 
			\STATE  $t$ $\gets$ $t+1$
			\STATE  $g_t$ $\gets$ $\bigtriangledown_\theta f_t(\theta_{t-1})$: Get gradients w.r.t stochastic objective at time step $t$
			\STATE  $m_t$ $\gets$ $\beta_1\cdot m_{t-1}+(1-\beta_1)\cdot g_t$: Update based first moment estimate
			\STATE  $v_t$ $\gets$ $\beta_2\cdot m_{t-1}+(1-\beta_2)\cdot (g_t)^2$: Update based second raw moment estimate
			\STATE  $\widehat{m}_t$ $\gets$ $m_t/(1-\beta_1^t)$: Compute bias-corrected first moment estimate
			\STATE  $\widehat{v}_t$ $\gets$ $v_t/(1-\beta_2^t)$: Compute bias-corrected second moment estimate
			\STATE  $\theta_t$ $\gets$ $\theta_{t-1} -\alpha\cdot\widehat{m}_t/(\sqrt{\widehat{v}_t}+\epsilon)$: Update parameter
			\ENDWHILE
			\RETURN $\theta_t$: Optimized parameters

		\end{algorithmic}    
	\end{algorithm}
	
	After deciding the input pattern, the next key step is to determine the structure of LSTM and the optimization algorithm for internal coefficients. Generally, the following parameters are selected with difficulties when building LSTM,
	\begin{enumerate}
		\item learning\_rate: the speed of reaching an optimal parameter value.
		\item num\_layers: the number of hidden layers.
		\item time\_steps: the number of unrolled steps. 
		\item batch\_size: the number of training samples for a mini-batch.
	\end{enumerate}
	
	Considering the relevant factors such as the depth of network learning, the complexity of network computing, the speed of calculation and so on, we learned the experience from literatures, then carried out a series of comparative experiments to adjust the above parameters.

	During the training process of LSTM, the adaptive moment estimation (Adam) algorithm~\cite{Adam} is applied to optimize the objective function. Adam is a first-order optimization algorithm which can replace the traditional stochastic gradient descent process. And, Adam can update neural network weights iteratively based on the training data as above steps.
	
	\subsection{Parameter optimization with PSO}
	The initialization of the input and output weights of LSTM is a random task, which however has a significant impact on the forecasting accuracy. Therefore, PSO is utilized to optimize these weights to improve the forecasting accuracy. Especially, the following steps are applied to each EMD component.

	\begin{algorithm}[H]  
		\caption{Particle Swarm Optimization} 
		\label{al1} 
		\begin{algorithmic}
			\renewcommand{\algorithmicrequire}{\textbf{Input:}}
			\REQUIRE  Initialize PSO with $m$ particles, $n$ iterations per particle.
			\FOR {each particle $P_i$}
			\STATE Initialize velocity $V_i$ and position $X_i$ 
			\STATE Evaluate particle $P_i$ and set $Pbest$ = $X_i$
			\ENDFOR
			\STATE $Gbest$ = min($Pbest$)
			\FOR {i = 1 to m}
			\FOR {j = 1 to n}
			\STATE Update the velocity and position of particle $P_i$
			\STATE Evaluate particle $P_i$
			\IF{fit($X_i$)$<$fit($Pbest$)}
			\STATE $Pbest$ = $X_i$
			\ENDIF
			\IF{fit($Pbest$)$<$fit($Gbest$)}
			\STATE $Gbest$ = $Pbest$
			\ENDIF
			\ENDFOR
			\ENDFOR
			\RETURN $Gbest$ and the corresponding $X_i$  		
		\end{algorithmic}    
	\end{algorithm}
	
	In the above steps, $V_i$ and $X_i$ represent the velocity and position of each particle. All particles search the optimal solution in the whole space, and record the answers as the current individual extremum, named personal best ($Pbest$). Comparing the individual extremum to all particles including itself in the whole particle swarm, the optimal individual extremum is found as the global best solution ($Gbest$)~\cite{PSO}.
	\subsection{Renormalization and linear addition}
	
	After training, a number of optimal LSTM neural networks corresponding to each EMD component are labeled by numerical values. Then the networks are used to predict the load of each part respectively, where the forecasting values are between 0 and 1. Thus, the results of each set are renormalized as follows,
	\begin{equation}
	X_{act}=X_{pre}\times(X_{max}-X_{min})+X_{min}\label{12}.
	\end{equation}
	In Equation (~\ref{12}), $X_{pre}$ represents the prediction value, $X_{max}$ and $X_{min}$ symbolize the maximum and minimum of each component. Finally, the prediction results are obtained by linearly adding all of the sets.
	
	\section{Experiments}
	In this study, the effectiveness of EMD-PSO-LSTM is exemplified by a realistic data set. In what follows, we first highlight the performance of EMD-PSO-LSTM. Then a series experiments were performed to analyze the impact of relevant factors such as input patterns and combination modes of EMD. Finally, comparisons among the proposed method and existing approaches are provided.
	
	\subsection{Data description}
	The data set is collected from the European Network of Excellence on Intelligent Technologies for Smart Adaptive Systems (EUNITE) competition. More specifically, the original data contains 8760 load values measured from 0:00 to 23:00 per day in year 1998.
	In this study, we use the first $335$ days as the training set to predict the 24 hour load of the $336^{th}$ day. Before application, these data is pre-processed through bad data cleaning, EMD decomposition and reorganization.
	\begin{figure}[htbp]
		
		\centering
		\includegraphics[width=9cm,height=12cm]{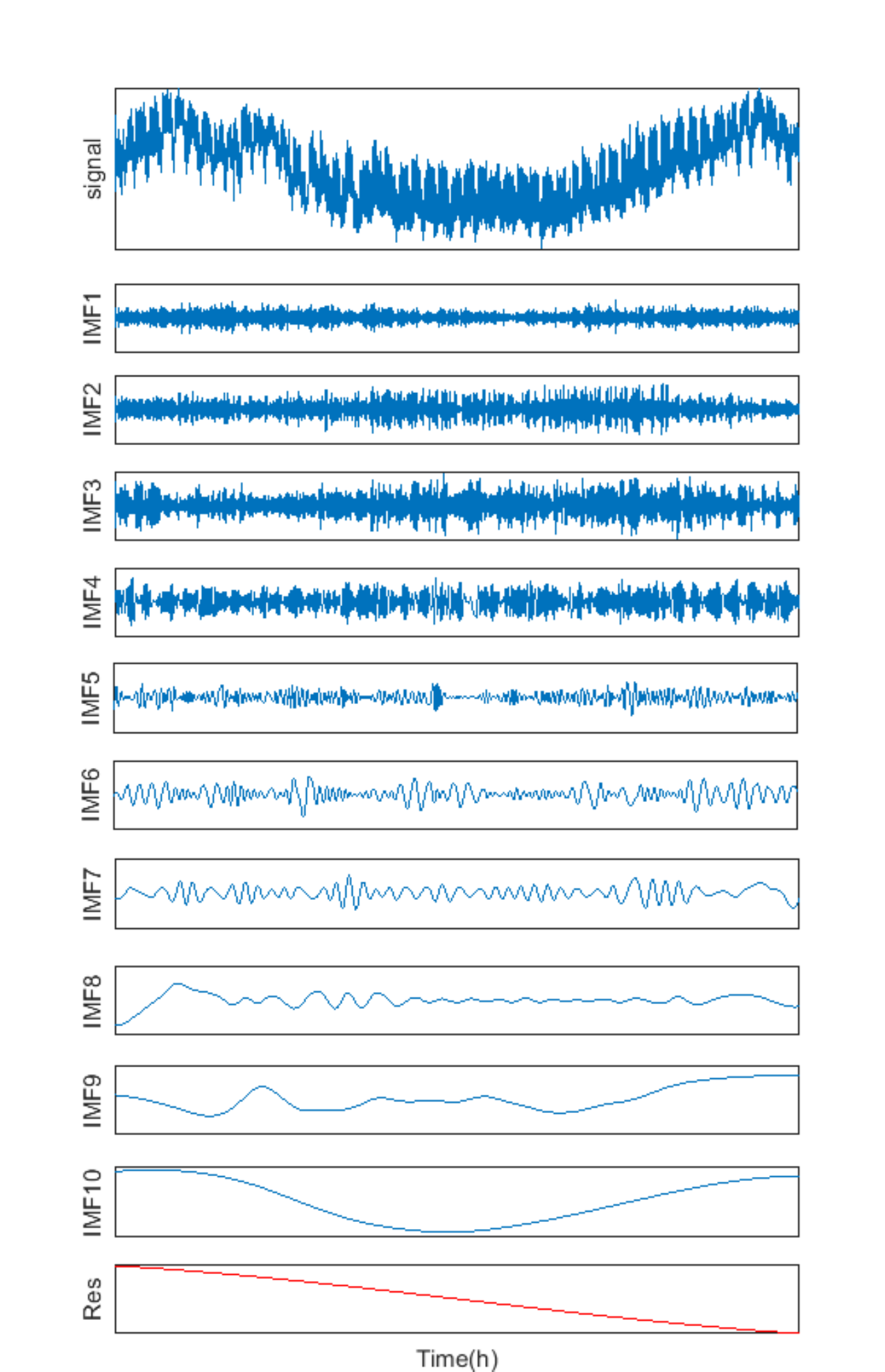}
		\caption{Load decomposition diagram by EMD}
		\label{fig:LSTM4}
	\end{figure}
	\subsection{Results and discussions}
	Here, mean absolute percentage error (MAPE) is applied to evaluate the performance of load forecasting,
	\begin{equation}
	MAPE=\frac{1}{n}\sum{\frac{\mid(X_{act}-X_{pre})\mid}{X_{act}}}\label{13},
	\end{equation}
	in Equation (~\ref{13}), $X_{act}$ represents the actual load data, $X_{pre}$ is the predictive value and $n$ equals 24.
	
	\subsubsection{Performance of EMD-PSO-LSTM}
	First, the training set is decomposed by EMD, where the raw data is divided into ten IMFs and one Res as shown in Fig. 5. Obviously, Fig. 5 shows that the change of components slows down gradually with the increase of the number of IMF. The reason is that, with the decreasing of the oscillating frequency, the IMF gradually appears like a oscillation mode similar to the sinusoidal signal, thus reducing the happening of random burst, which facilitates the networks to learn the local features presented by each component.
	
	Then the proposed method is used to predict the load data of the $336^{th}$ day, which is based on MIX3, i.e. IMF 1 to 3 are superimposed to form the high frequency component, and the remaining IMFs and Res are treated as separate components. The results are depicted in Fig. 6 and Fig. 7.
	
	\begin{figure}[H]
		\begin{minipage}[h]{0.5\linewidth}
			\centering
			\includegraphics[height=5.5cm,width=7.5cm]{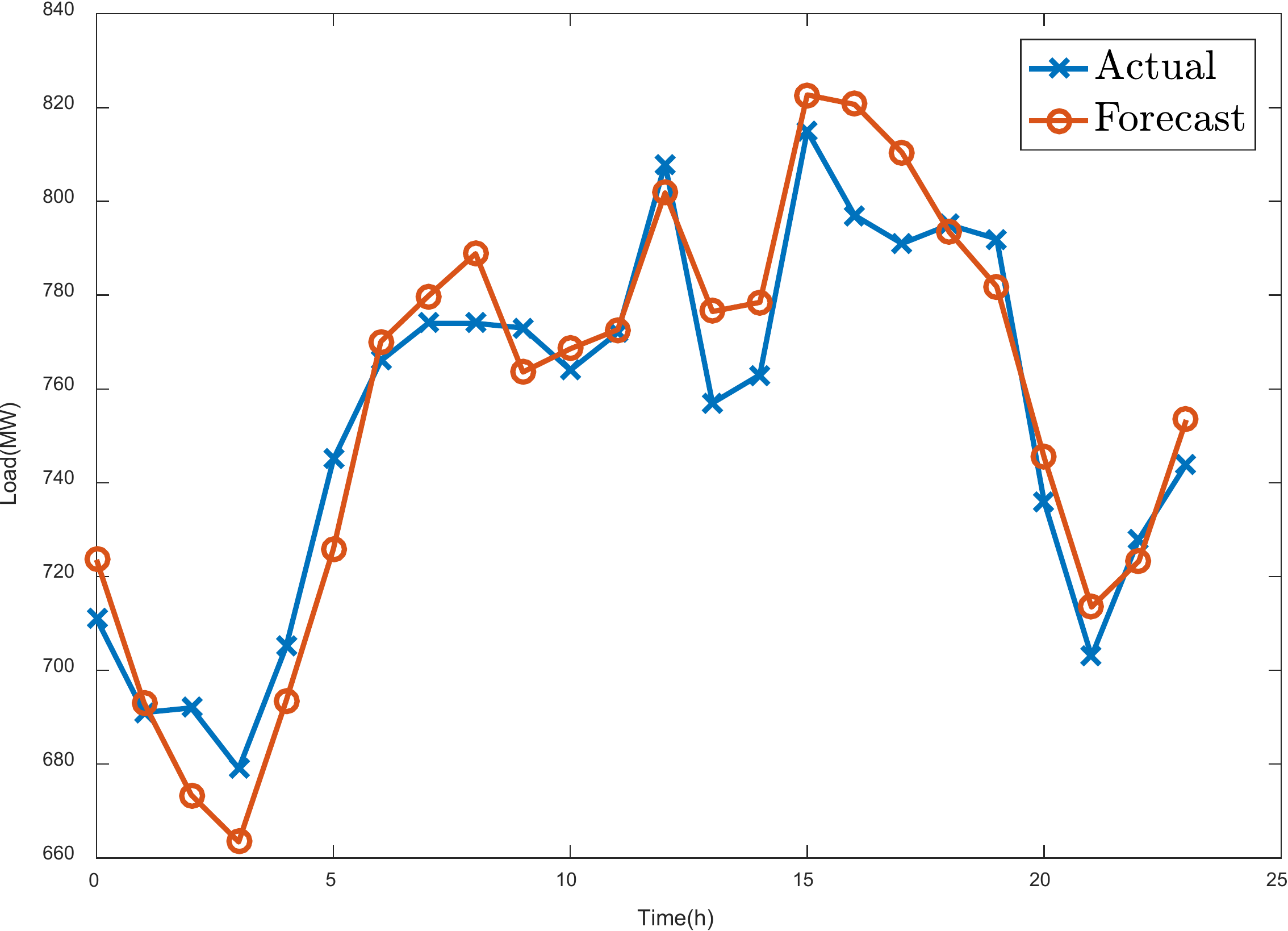}
			\caption{Performance of EMD-PSO-LSTM}
			\label{fig:LSTM05}
		\end{minipage}%
		\hfill
		\begin{minipage}[h]{0.5\linewidth}
			\centering
			\includegraphics[height=5.5cm,width=7.5cm]{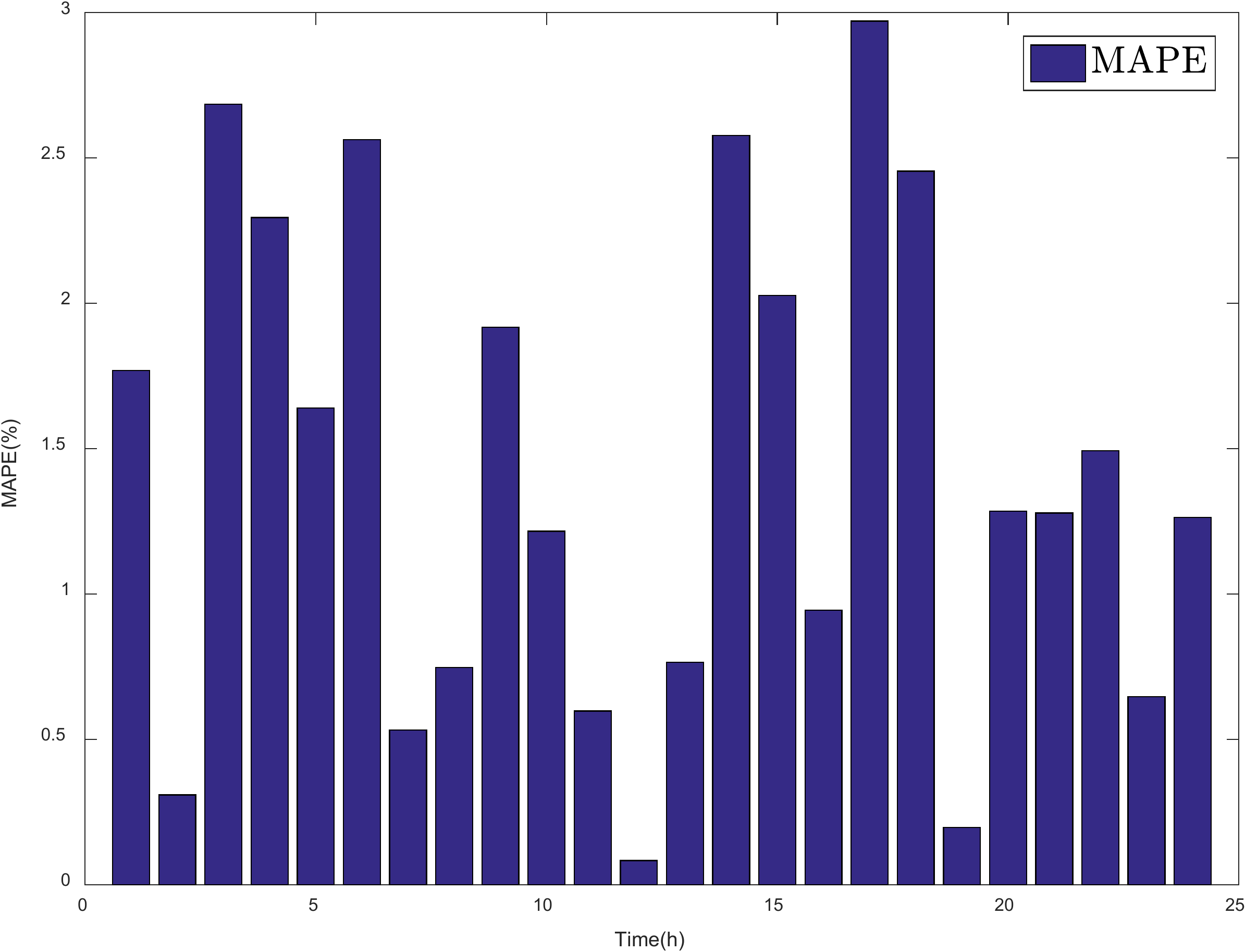}
			\caption{MAPE of EMD-PSO-LSTM}
			\label{fig:LSTM06}
		\end{minipage}
		
	\end{figure}

	Fig. 6 shows that the predicted load curve is close to the actual one. Besides, Fig. 7 demonstrates that the maximum,  minimum and mean forecasting errors are 2.9708\%, 0.0833\% and 1.4274\%, while most of errors are below 1.5\%. Therefore, it can be concluded that the proposed method is effective.
	
	\subsubsection{Impact of input pattern}
	
	\begin{figure}[h]
		
		\centering
		\includegraphics[height=4.8cm,width=15cm]{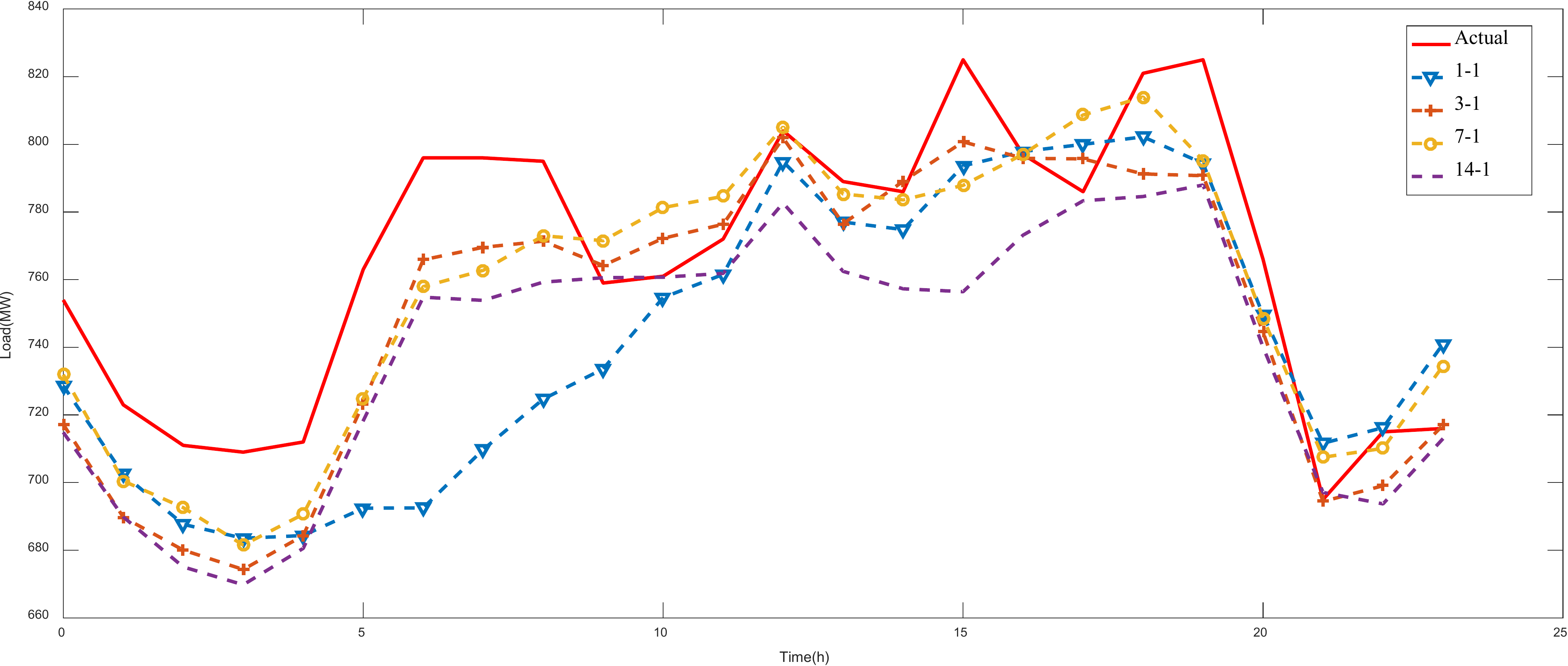}
		\caption{Experimental results of 1-1,3-1,7-1,14-1 input patterns}
		\label{fig:LSTM5}
	\end{figure}
	The input pattern is composed of the characteristics and labels of the training data. Generally, an appropriate input pattern can accelerate LSTM to learn the features of the training data and get better prediction results. In order to evaluate the effects of different input patterns on prediction results, a series of experiments were performed. The results are described in Fig. 8 and Table ~\ref{table I}, where 1-1, 3-1 represent the $N$-to-one input pattern as explained in Subsection III B.
	
	\renewcommand\arraystretch{0.6}
	\begin{table}[htbp]
		\small
		\setlength{\tabcolsep}{0.35pt}
		\caption{\label{table I}  MAPE of input pattern experiments}
		\begin{center}
			\begin{tabular}{c|cc|cc|cc|cc}\toprule
				Actual load & 1-1 & MAPE & 3-1 & MAPE & 7-1 & MAPE & 14-1 & MAPE \\
				\hline
				754.00 & 728.63 & 3.36 & 717.23 & 4.88 & 731.91 & 2.93 & 714.86 & 5.19 \\
				723.00 & 702.50 & 2.83 & 689.78 & 4.59 & 700.49 & 3.11 & 689.85 & 4.59 \\
				711.00 & 687.66 & 3.28 & 680.04 & 4.35 & 692.59 & 2.59 & 675.07 & 5.05 \\
				709.00 & 683.53 & 3.59 & 674.31 & 4.89 & 681.53 & 3.87 & 669.79 & 5.53 \\
				712.00 & 684.34 & 3.88 & 684.33 & 3.89 & 690.63 & 3.00 & 680.56 & 4.42 \\
				763.00 & 692.43 & 9.25 & 723.36 & 5.20 & 724.92 & 4.99 & 718.27 & 5.86 \\
				796.00 & 692.52 & 13.00 & 765.87 & 3.79 & 757.92 & 4.78 & 754.81 & 5.17 \\
				796.00 & 709.76 & 10.83 & 769.57 & 3.32 & 762.75 & 4.18 & 753.87 & 5.29 \\
				795.00 & 724.68 & 8.85 & 771.37 & 2.97 & 772.96 & 2.77 & 759.27 & 4.49 \\
				759.00 & 733.62 & 3.34 & 764.19 & 0.68 & 771.53 & 1.65 & 760.56 & 0.21 \\
				761.00 & 754.71 & 0.83 & 772.23 & 1.48 & 781.30 & 2.67 & 760.71 & 0.04 \\
				772.00 & 761.55 & 1.35 & 776.37 & 0.57 & 784.64 & 1.64 & 761.85 & 1.31 \\
				804.00 & 794.87 & 1.14 & 801.88 & 0.26 & 805.13 & 0.14 & 782.64 & 2.66 \\
				789.00 & 777.03 & 1.52 & 776.51 & 1.58 & 785.32 & 0.47 & 762.45 & 3.36 \\
				786.00 & 774.80 & 1.43 & 789.07 & 0.39 & 783.59 & 0.31 & 757.29 & 3.65 \\
				825.00 & 793.63 & 3.80 & 800.75 & 2.94 & 787.89 & 4.50 & 756.44 & 8.31 \\
				797.00 & 797.90 & 0.11 & 795.88 & 0.14 & 797.08 & 0.01 & 773.17 & 2.99 \\
				786.00 & 799.99 & 1.78 & 795.70 & 1.23 & 808.67 & 2.88 & 783.31 & 0.34 \\
				821.00 & 802.22 & 2.29 & 791.25 & 3.62 & 813.99 & 0.85 & 784.58 & 4.44 \\
				825.00 & 794.36 & 3.71 & 790.73 & 4.15 & 795.23 & 3.61 & 787.98 & 4.49 \\
				766.00 & 749.53 & 2.15 & 744.51 & 2.81 & 748.66 & 2.26 & 740.06 & 3.39 \\
				695.00 & 711.62 & 2.39 & 694.43 & 0.08 & 707.49 & 1.80 & 697.09 & 0.30 \\
				715.00 & 716.24 & 0.17 & 699.06 & 2.23 & 710.12 & 0.68 & 693.71 & 2.98 \\
				716.00 & 741.00 & 3.49 & 716.97 & 0.13 & 734.51 & 2.59 & 713.00 & 0.42 \\
				\hline
				Average MAPE&        & 3.68 & 	&2.51 & 	&2.43 & 	&3.52 \\
				\bottomrule
			\end{tabular}
		\end{center}
	\end{table} 
	Clearly, Fig. 8 and Table ~\ref{table I} depict that the maximum MAPEs of each pattern are 13.00\%, 5.20\%, 4.99\%, 8.31\%, the minimum MAPEs of each pattern are 0.11\%, 0.08\%, 0.01\%, 0.04\%, and the mean MAPEs of each pattern are 3.68\%, 2.51\%, 2.43\%, 3.52\% separately. In general, as the $N$ value rises, the prediction accuracy increases first and then decreases. Since the load shows weekly cyclical, it can be speculated that the forecast result with the seven-to-one input pattern is more accurate than the others. According to the above analysis, the seven-to-one input pattern is adopted in all subsequent experiments.
	\subsubsection{Effectiveness of IMFs and Res combinations}
	As mentioned in Subsection II B, the raw data is decomposed by EMD, and each component shows a short-correlation based on the frequency.
	Restructuring the components with similar frequency not only preserves the main feature information of load data, but also reduces the number of prediction models, which improves the prediction efficiency and lightens the computation burden.
	In order to evaluate the effects of different IMF recombinations on the forecast accuracy, a number of experiments were carried out. In the experiments, MIX$n$ indicates that IMF 1 to $n$ are stacked as a high frequency component, the rest IMFs and Res are treated as a low frequency component. 
	The results are depicted in Fig. 9 and Table ~\ref{table II}.
	
	Fig. 9 and Table ~\ref{table II} show that the mean MAPEs of each combination are 2.08\%, 2.13\%, 1.55\%, 1.70\%, 1.62\%, 3.20\%, 2.11\%, 3.71\%, 3.12\%, 3.94\% and 3.51\%. Obviously, MIX3 which combines IMF 1 to 3 into a high frequency component, has obtained better forecasting results than other combinations. The main reason could be that the reorganization of IMF 1 to 3 reduces certain errors of high frequency parts. In summary, MIX3 is the most effective combination in this data set.
	\renewcommand\arraystretch{0.9}
	\begin{sidewaystable*}[p]
		\small
		\setlength{\tabcolsep}{0.35pt}
		\caption{\label{table II}  MAPE of IMFs and RES combinations}
		\begin{center}
			\begin{tabular}{c|cc|cc|cc|cc|cc|cc|cc|cc|cc|cc|cc}\toprule
				Actual load & MIX1 & MAPE & MIX2 & MAPE & MIX3 & MAPE & MIX4 & MAPE & MIX5 & MAPE & MIX6 & MAPE & MIX7 & MAPE & MIX8 & MAPE & MIX9 & MAPE & MIX10 & MAPE & MIX11 & MAPE\\
				\hline
				705.00 & 694.91 & 1.43 & 680.05 & 3.54 & 689.86 & 2.15 & 696.29 & 1.24 & 702.29 & 0.38 & 702.63 & 0.34 & 695.82 & 1.30 & 697.98 & 1.00 & 694.47 & 1.49 & 697.68 & 1.04 & 705.23 & 0.03 \\
				665.00 & 671.70 & 1.01 & 663.61 & 0.21 & 667.43 & 0.36 & 672.14 & 1.07 & 682.43 & 2.62 & 674.36 & 1.41 & 671.99 & 1.05 & 676.12 & 1.67 & 664.89 & 0.02 & 664.32 & 0.10 & 672.41 & 1.11 \\
				669.00 & 655.59 & 2.01 & 639.95 & 4.34 & 645.62 & 3.50 & 656.38 & 1.89 & 661.94 & 1.06 & 656.02 & 1.94 & 653.70 & 2.29 & 660.33 & 1.30 & 649.73 & 2.88 & 641.41 & 4.12 & 662.82 & 0.92 \\
				664.00 & 659.12 & 0.73 & 629.76 & 5.16 & 642.57 & 3.23 & 652.31 & 1.76 & 659.69 & 0.65 & 654.33 & 1.46 & 648.26 & 2.37 & 652.97 & 1.66 & 642.83 & 3.19 & 642.12 & 3.29 & 647.79 & 2.44 \\
				677.00 & 647.75 & 4.32 & 648.14 & 4.26 & 658.38 & 2.75 & 661.32 & 2.32 & 664.58 & 1.84 & 666.17 & 1.60 & 654.89 & 3.27 & 659.57 & 2.58 & 650.07 & 3.98 & 651.12 & 3.82 & 660.65 & 2.42 \\
				696.00 & 714.87 & 2.71 & 697.45 & 0.21 & 698.81 & 0.40 & 702.03 & 0.87 & 709.24 & 1.90 & 712.04 & 2.31 & 700.30 & 0.62 & 700.97 & 0.71 & 701.54 & 0.80 & 692.10 & 0.56 & 714.37 & 2.64 \\
				753.00 & 768.81 & 2.10 & 758.98 & 0.79 & 744.55 & 1.12 & 745.21 & 1.03 & 745.87 & 0.95 & 746.83 & 0.82 & 744.08 & 1.18 & 738.33 & 1.95 & 741.69 & 1.50 & 717.18 & 4.76 & 756.61 & 0.48 \\
				751.00 & 764.69 & 1.82 & 767.07 & 2.14 & 761.91 & 1.45 & 752.34 & 0.18 & 756.78 & 0.77 & 741.51 & 1.26 & 750.35 & 0.09 & 740.86 & 1.35 & 742.42 & 1.14 & 721.62 & 3.91 & 747.99 & 0.40 \\
				785.00 & 775.03 & 1.27 & 767.88 & 2.18 & 768.51 & 2.10 & 762.15 & 2.91 & 769.42 & 1.99 & 748.61 & 4.64 & 773.46 & 1.47 & 750.73 & 4.37 & 748.66 & 4.63 & 733.99 & 6.50 & 749.35 & 4.54 \\
				772.00 & 760.48 & 1.49 & 771.64 & 0.05 & 775.83 & 0.50 & 766.09 & 0.77 & 773.37 & 0.18 & 744.11 & 3.61 & 768.63 & 0.44 & 749.29 & 2.94 & 744.40 & 3.57 & 740.33 & 4.10 & 749.17 & 2.96 \\
				785.00 & 732.81 & 6.65 & 775.50 & 1.21 & 782.15 & 0.36 & 768.36 & 2.12 & 776.11 & 1.13 & 744.09 & 5.21 & 776.27 & 1.11 & 746.66 & 4.88 & 751.22 & 4.30 & 751.62 & 4.25 & 749.08 & 4.58 \\
				775.00 & 767.00 & 1.03 & 782.80 & 1.01 & 782.15 & 0.92 & 767.74 & 0.94 & 768.94 & 0.78 & 745.62 & 3.79 & 758.78 & 2.09 & 743.26 & 4.09 & 748.00 & 3.48 & 744.88 & 3.89 & 744.56 & 3.93 \\
				801.00 & 797.26 & 0.47 & 805.70 & 0.59 & 799.70 & 0.16 & 788.35 & 1.58 & 779.60 & 2.67 & 770.34 & 3.83 & 794.51 & 0.81 & 756.42 & 5.57 & 774.69 & 3.28 & 771.22 & 3.72 & 774.76 & 3.28 \\
				791.00 & 771.51 & 2.46 & 794.83 & 0.48 & 785.10 & 0.75 & 775.44 & 1.97 & 764.83 & 3.31 & 748.87 & 5.33 & 774.27 & 2.11 & 738.22 & 6.67 & 749.58 & 5.24 & 746.39 & 5.64 & 744.01 & 5.94 \\
				787.00 & 784.73 & 0.29 & 797.36 & 1.32 & 784.74 & 0.29 & 767.79 & 2.44 & 762.20 & 3.15 & 744.78 & 5.36 & 759.38 & 3.51 & 734.05 & 6.73 & 753.41 & 4.27 & 737.95 & 6.23 & 732.04 & 6.98 \\
				829.00 & 778.04 & 6.15 & 785.96 & 5.19 & 790.13 & 4.69 & 773.37 & 6.71 & 768.59 & 7.29 & 743.65 & 10.30 & 776.01 & 6.39 & 729.47 & 12.01 & 751.70 & 9.32 & 732.04 & 11.70 & 739.64 & 10.78\\ 
				805.00 & 802.41 & 0.32 & 791.61 & 1.66 & 815.28 & 1.28 & 796.26 & 1.09 & 792.09 & 1.60 & 766.48 & 4.78 & 807.55 & 0.32 & 740.13 & 8.06 & 774.03 & 3.85 & 774.54 & 3.78 & 748.02 & 7.08 \\
				819.00 & 791.96 & 3.30 & 801.18 & 2.18 & 810.27 & 1.07 & 814.21 & 0.58 & 807.98 & 1.35 & 795.26 & 2.90 & 836.41 & 2.13 & 757.82 & 7.47 & 796.45 & 2.75 & 777.10 & 5.12 & 774.57 & 5.43 \\
				803.00 & 825.50 & 2.80 & 825.29 & 2.78 & 822.05 & 2.37 & 814.55 & 1.44 & 807.77 & 0.59 & 806.82 & 0.48 & 830.34 & 3.40 & 782.79 & 2.52 & 803.66 & 0.08 & 805.96 & 0.37 & 785.69 & 2.16 \\
				793.00 & 801.97 & 1.13 & 824.00 & 3.91 & 825.12 & 4.05 & 805.34 & 1.56 & 797.62 & 0.58 & 777.28 & 1.98 & 808.34 & 1.93 & 785.59 & 0.93 & 783.06 & 1.25 & 774.38 & 2.35 & 774.48 & 2.34 \\
				749.00 & 778.54 & 3.94 & 759.45 & 1.40 & 762.10 & 1.75 & 759.88 & 1.45 & 762.24 & 1.77 & 727.92 & 2.81 & 731.74 & 2.30 & 740.18 & 1.18 & 728.47 & 2.74 & 724.51 & 3.27 & 720.48 & 3.81 \\
				717.00 & 715.34 & 0.23 & 704.08 & 1.80 & 711.35 & 0.79 & 725.98 & 1.25 & 709.80 & 1.00 & 678.06 & 5.43 & 681.87 & 4.90 & 690.52 & 3.69 & 682.06 & 4.87 & 681.84 & 4.90 & 684.46 & 4.54 \\
				707.00 & 714.91 & 1.12 & 693.47 & 1.91 & 700.18 & 0.96 & 714.60 & 1.08 & 701.58 & 0.77 & 673.10 & 4.80 & 678.41 & 4.04 & 684.98 & 3.11 & 675.77 & 4.42 & 665.59 & 5.86 & 679.44 & 3.90 \\
				710.00 & 718.23 & 1.16 & 689.20 & 2.93 & 711.15 & 0.16 & 728.05 & 2.54 & 713.31 & 0.47 & 706.17 & 0.54 & 700.11 & 1.39 & 691.53 & 2.60 & 696.73 & 1.87 & 701.68 & 1.17 & 699.21 & 1.52 \\
				
				\hline
				Average MAPE&     &	2.08 &   & 2.13 &      &1.55 &      &1.70 &      & 1.62 &      & 3.20 &      & 2.11 &      & 3.71 &      & 3.12 &      & 3.94 &      & 3.51 \\
				\bottomrule
			\end{tabular}
		\end{center}
	\end{sidewaystable*}
	\begin{figure}[H]
		
		\centering
		\includegraphics[height=4.8cm,width=15cm]{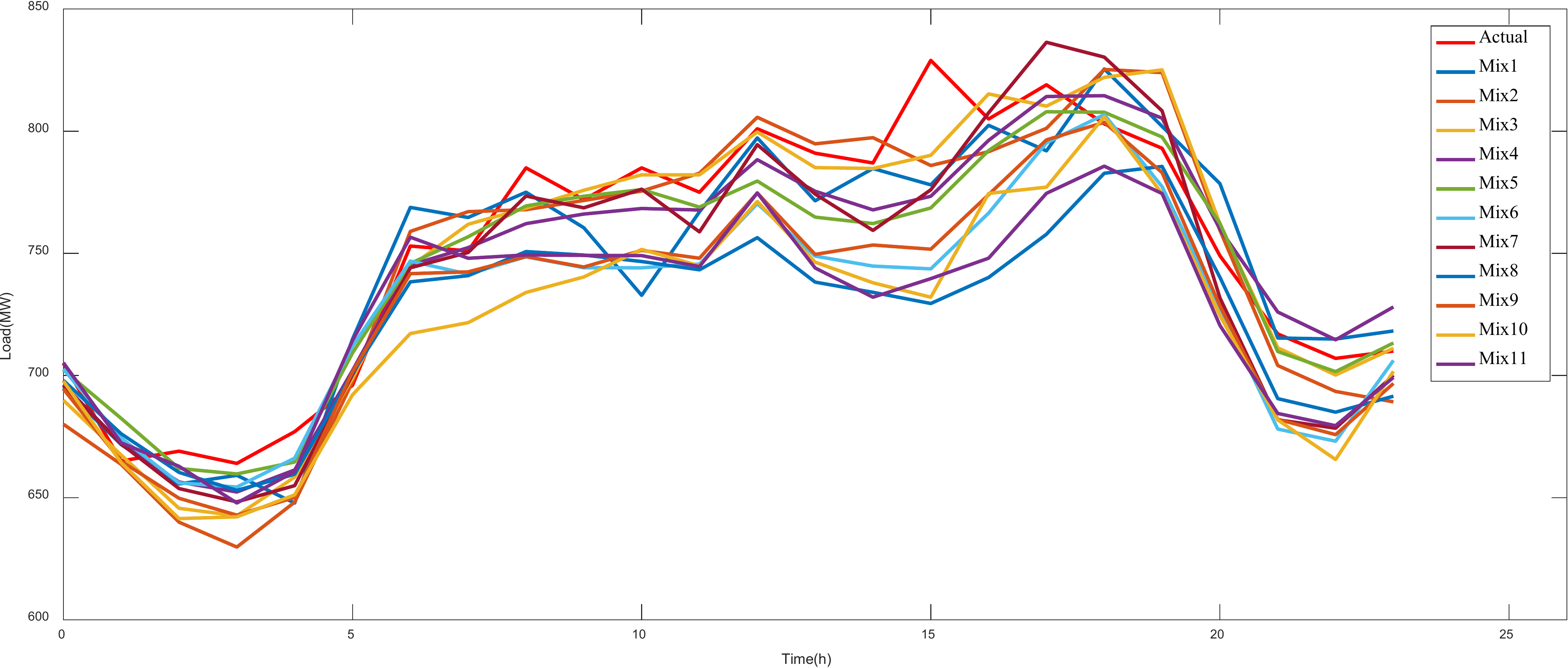}
		\caption{Combination experiment results of IMFs and RES}
		\label{fig:LSTM6}
	\end{figure}
	
	\subsubsection{Comparisons of LSTM, EMD-LSTM, PSO-LSTM, RNN, GRU and EMD-PSO-LSTM}
	\begin{figure}[H]
		
		\centering
		\includegraphics[width=15cm,height=9cm]{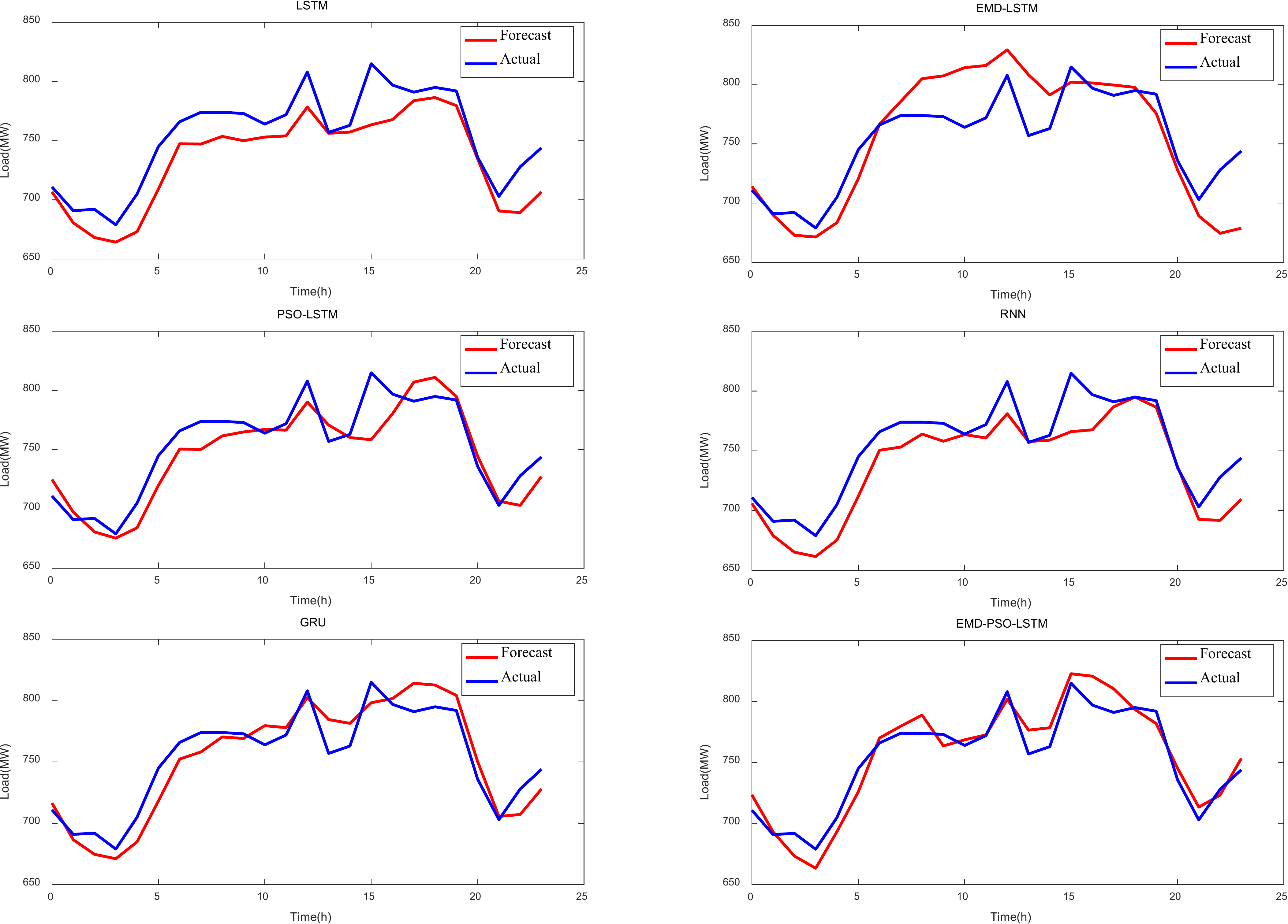}
		\caption{Experimental results of LSTM, EMD-LSTM, PSO-LSTM, RNN, GRU, EMD-PSO-LSTM}
		\label{fig:LSTM7}
	\end{figure}
	
	In this paper, EMD-PSO-LSTM is used to predict the load of the $336^{th}$ day. Meanwhile, other methods are utilized to prove the validity of the proposed method. In addition to the relevant factors mentioned above, the other parameters of all methods are set as follows.
	
	\begin{itemize}	
		\item[] $RNN$ $network$ $layer$ $number$ = $1$ 
		\item[] $Hidden$ $neuron$ $number$ = $10$
		\item[] $Training$ $batch$ $size$ = $64$
		\item[] $Learning$ $rate$ = $0.005$
		\item[] $Training$ $method$ = $Adamoptimizer$
		\item[] $Loss$ $feedback$ = $RMSE$
		
	\end{itemize}
	\begin{equation}
	RMSE=\sqrt{\frac{\sum{({X_{a,b}-Mean})^2}}{n}}\label{14}.
	\end{equation}
	In order to make a more comprehensive comparison of the effectiveness of the proposed method, experiments are carried out from two aspects. First, EMD-PSO-LSTM is compared with the original LSTM, LSTM with EMD, and LSTM with PSO to describe the process of improvement. Second, the proposed method is compared with existing approaches such as RNN~\cite{RNN4} and GRU~\cite{GRU1} to further confirm the superiority of EMD-PSO-LSTM.
	Fig. 10 shows the actual 24 hour load of the $336^{th}$ day and the predicted values obtained by each method.
	Obviously, the forecast load curve of EMD-PSO-LSTM is the closest to the actual one. 
	In order to describe the prediction accuracy of each method more intuitively, the MAPE of 24 hour points is shown in Fig. 11 and Table ~\ref{table III}. The mean MAPEs of each method are 2.61\%, 2.97\%, 1.89\%, 2.21\%, 1.77\% and 1.43\%, among which EMD-PSO-LSTM has the smallest error.
	\begin{figure}[h]
		
		\centering
		\includegraphics[width=15cm,height=4.8cm]{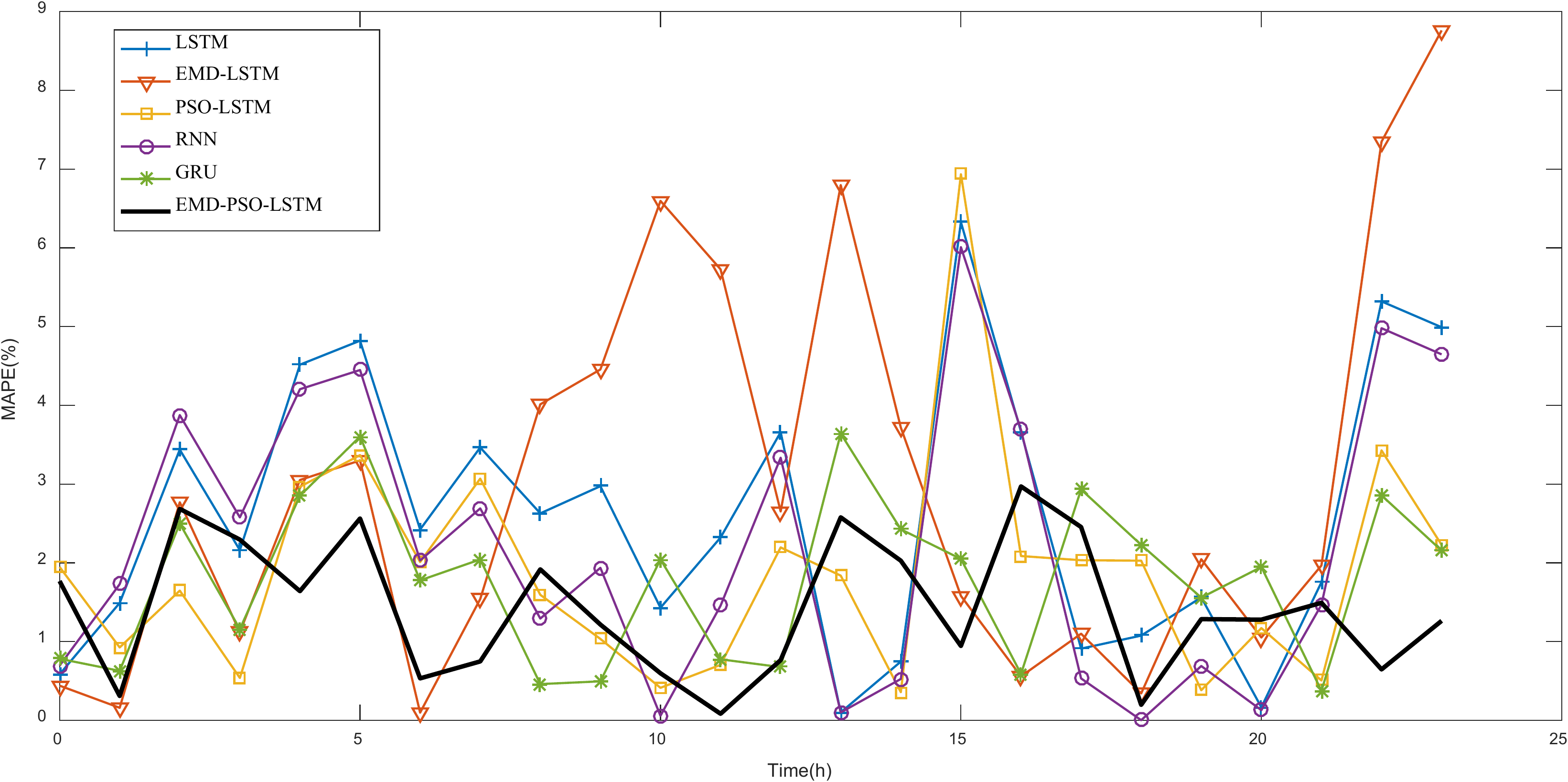}
		\caption{MAPE of LSTM, EMD-LSTM, PSO-LSTM, RNN, GRU, EMD-PSO-LSTM}
		\label{fig:LSTM8}
	\end{figure}
	
	\renewcommand\arraystretch{0.5}
	\begin{table}[h]
		\small
		\setlength{\tabcolsep}{0.35pt}
		\caption{\label{table III}  MAPE of LSTM, EMD-LSTM, PSO-LSTM, RNN, GRU, EMD-PSO-LSTM}
		\begin{center}
			\begin{tabular}{c|cc|cc|cc|cc|cc|cc}\toprule
				Actual load & LSTM  & MAPE  & EMD-LSTM  & MAPE  & PSO-LSTM & MAPE & RNN & MAPE & GRU & MAPE & EMD-PSO-LSTM & MAPE \\
				\hline
				711.00 & 706.85 & 0.58 & 714.13 & 0.44 & 724.85 & 1.95 & 706.06 & 0.69 & 716.56 & 0.78 & 723.57 & 1.77 \\
				691.00 & 680.70 & 1.49 & 689.90 & 0.16 & 697.40 & 0.93 & 679.05 & 1.73 & 686.66 & 0.63 & 693.13 & 0.31 \\
				692.00 & 668.21 & 3.44 & 672.84 & 2.77 & 680.60 & 1.65 & 665.18 & 3.88 & 674.71 & 2.50 & 673.42 & 2.68 \\
				679.00 & 664.33 & 2.16 & 671.40 & 1.12 & 675.34 & 0.54 & 661.50 & 2.58 & 671.13 & 1.16 & 663.42 & 2.30 \\
				705.00 & 673.16 & 4.52 & 683.48 & 3.05 & 684.09 & 2.97 & 675.35 & 4.21 & 684.90 & 2.85 & 693.44 & 1.64 \\
				745.00 & 709.08 & 4.82 & 720.41 & 3.30 & 719.91 & 3.37 & 711.86 & 4.45 & 718.15 & 3.60 & 725.91 & 2.56 \\
				766.00 & 747.43 & 2.42 & 766.75 & 0.10 & 750.60 & 2.01 & 750.45 & 2.03 & 752.38 & 1.78 & 770.08 & 0.53 \\
				774.00 & 747.18 & 3.46 & 785.92 & 1.54 & 750.24 & 3.07 & 753.17 & 2.69 & 758.19 & 2.04 & 779.79 & 0.75 \\
				774.00 & 753.66 & 2.63 & 805.06 & 4.01 & 761.66 & 1.59 & 764.00 & 1.29 & 770.44 & 0.46 & 788.84 & 1.92 \\
				773.00 & 750.01 & 2.97 & 807.44 & 4.46 & 765.01 & 1.03 & 758.03 & 1.94 & 769.17 & 0.50 & 763.60 & 1.22 \\
				764.00 & 753.07 & 1.43 & 814.39 & 6.60 & 767.15 & 0.41 & 763.56 & 0.06 & 779.61 & 2.04 & 768.57 & 0.60 \\
				772.00 & 754.08 & 2.32 & 816.22 & 5.73 & 766.56 & 0.70 & 760.76 & 1.46 & 777.98 & 0.77 & 772.64 & 0.08 \\
				808.00 & 778.43 & 3.66 & 829.43 & 2.65 & 790.26 & 2.20 & 781.06 & 3.33 & 802.54 & 0.68 & 801.82 & 0.77 \\
				757.00 & 756.23 & 0.10 & 808.45 & 6.80 & 770.91 & 1.84 & 757.76 & 0.10 & 784.56 & 3.64 & 776.51 & 2.58 \\
				763.00 & 757.31 & 0.75 & 791.37 & 3.72 & 760.27 & 0.36 & 758.99 & 0.53 & 781.51 & 2.43 & 778.47 & 2.03 \\
				815.00 & 763.44 & 6.33 & 802.12 & 1.58 & 758.48 & 6.93 & 766.03 & 6.01 & 798.30 & 2.05 & 822.70 & 0.94 \\
				797.00 & 767.79 & 3.66 & 801.45 & 0.56 & 780.37 & 2.09 & 767.60 & 3.69 & 801.70 & 0.59 & 820.68 & 2.97 \\
				791.00 & 783.80 & 0.91 & 799.70 & 1.10 & 807.08 & 2.03 & 786.79 & 0.53 & 814.19 & 2.93 & 810.42 & 2.45 \\
				795.00 & 786.45 & 1.08 & 797.72 & 0.34 & 811.12 & 2.03 & 795.02 & 0.00 & 812.71 & 2.23 & 793.43 & 0.20 \\
				792.00 & 779.63 & 1.56 & 775.80 & 2.05 & 795.03 & 0.38 & 786.54 & 0.69 & 804.35 & 1.56 & 781.82 & 1.29 \\
				736.00 & 734.78 & 0.17 & 728.26 & 1.05 & 744.62 & 1.17 & 736.95 & 0.13 & 750.33 & 1.95 & 745.41 & 1.28 \\
				703.00 & 690.67 & 1.75 & 689.17 & 1.97 & 706.56 & 0.51 & 692.69 & 1.47 & 705.63 & 0.37 & 713.49 & 1.49 \\
				728.00 & 689.27 & 5.32 & 674.50 & 7.35 & 703.05 & 3.43 & 691.75 & 4.98 & 707.20 & 2.86 & 723.29 & 0.65 \\
				744.00 & 706.86 & 4.99 & 678.81 & 8.76 & 727.43 & 2.23 & 709.42 & 4.65 & 727.97 & 2.16 & 753.40 & 1.26 \\
				\hline
				Average  MAPE&	&2.61 & 	&2.97 & 	&1.89 & 	&2.21 & 	&1.77 & 	&1.43 \\
				\bottomrule
			\end{tabular}

		\end{center}
	\end{table}
	Objectively, in EMD-LSTM, because of the irregular variation of high frequency time series, it is difficult to understand the trend of small fluctuations, which leads to high errors of the whole prediction. In PSO-LSTM, although PSO improves the initialization of LSTM parameters, the method does not capture internal characteristics of load data, which leads to the inaccuracy of the overall trend. However, as for EMD-PSO-LSTM, it combines the advantage of EMD and PSO, which leads the errors of all points below 3\% and improves the forecasting accuracy evidently. 
	Moreover, compared with RNN and GRU, EMD-PSO-LSTM has precision advantages of nearly 0.8\% and 0.4\% separately. 
	
	Based on the above knowledge, it can be concluded that the proposed method is effective for STLF.
	
	\section{Conclusion}
	
	In this paper, a novel EMD-PSO-LSTM method for STLF was proposed, which utilizes load data as the feature to predict the 24 hour load. Based on a series of experiments, EMD-PSO-LSTM is proved to be effective. Especially, the prediction accuracy of EMD-PSO-LSTM has reached 98.5726\%, where the errors of all points are below 3\%, and the minimum error is 0.0833\%. Compared with traditional methods, EMD-PSO-LSTM can learn the internal characteristics of the load data more detailedly thus improving the prediction accuracy by nearly 1.2\%. Besides, compared with the state of the art methods, EMD-PSO-LSTM also outperforms them by nearly 0.4\%. 
	
	In the future, the following aspects could be proceeded,
	\begin{enumerate}
		\item The parameter optimization method could be further investigated to both improve the forecast accuracy and reduce the computational burden of the proposed method.
		\item The effect of other factors such as the impact of unexpected or major events on power load could be taken into account, thus improving the prediction accuracy and expanding the application space of the forecasting method.
	\end{enumerate}
	
	\section*{Acknowledgment}
	
	This work was supported by the National Natural Science Foundation of China (Grant No. 61603176 ), the Natural Science Foundation of Jiangsu Province (Grant No. BK20160632 ), and the Fundamental Research Funds for the Central Universities.
	
	\section*{Reference}

	\bibliographystyle{unsrt}
	\bibliography{document}

\begin{thebibliography}{10}

\bibitem{Intro1}
Ergun Yukseltan, Ahmet Yucekaya, and Ayse~Humeyra Bilge.
\newblock Forecasting electricity demand for turkey: Modeling periodic
  variations and demand segregation.
\newblock {\em Applied Energy}, 193:287--296, 2017.

\bibitem{Mean}
Bo~Wang, Shuming Wang, Xian~Zhong Zhou, and Junzo Watada.
\newblock Two-stage multi-objective unit commitment optimization under hybrid
  uncertainties.
\newblock In {\em International Conference on Genetic \& Evolutionary
  Computing}, pages 128--131, 2016.

\bibitem{Mean2}
Kody~M. Powell, Akshay Sriprasad, Wesley~J. Cole, and Thomas~F. Edgar.
\newblock Heating, cooling, and electrical load forecasting for a large-scale
  district energy system.
\newblock {\em Energy}, 74(5):877--885, 2014.

\bibitem{Intro3}
M.~R. Alrashidi and K.~M. El-Naggar.
\newblock Long term electric load forecasting based on particle swarm
  optimization.
\newblock {\em Applied Energy}, 87(1):320--326, 2010.

\bibitem{Intro4}
Matteo~De Felice, Andrea Alessandri, and Franco Catalano.
\newblock Seasonal climate forecasts for medium-term electricity demand
  forecasting.
\newblock {\em Applied Energy}, 137:435--444, 2015.

\bibitem{Intro5}
M.~Moazzami, A.~Khodabakhshian, and R.~Hooshmand.
\newblock A new hybrid day-ahead peak load forecasting method for iran’s
  national grid.
\newblock {\em Applied Energy}, 101(C):489--501, 2013.

\bibitem{ARIMA1}
Cheng~Ming Lee and Chia~Nan Ko.
\newblock Short-term load forecasting using lifting scheme and arima models.
\newblock {\em Expert Systems with Applications}, 38(5):5902--5911, 2011.

\bibitem{KF}
Che Guan, Peter~B. Luh, Laurent~D. Michel, and Zhiyi Chi.
\newblock Hybrid kalman filters for very short-term load forecasting and
  prediction interval estimation.
\newblock {\em IEEE Transactions on Power Systems}, 28(4):3806--3817, 2013.

\bibitem{AR1}
D.~H. Vu, K.~M. Muttaqi, A.~P. Agalgaonkar, A.~Bouzerdoum, D.~H. Vu, K.~M.
  Muttaqi, A.~P. Agalgaonkar, and A.~Bouzerdoum.
\newblock Short-term electricity demand forecasting using autoregressive based
  time varying model incorporating representative data adjustment.
\newblock {\em Applied Energy}, 205:790--801, 2017.

\bibitem{ELM1}
Song Li, Lalit Goel, and Peng Wang.
\newblock An ensemble approach for short-term load forecasting by extreme
  learning machine.
\newblock {\em Applied Energy}, 170:22--29, 2016.

\bibitem{GRNN1}
Liye Xiao, Wei Shao, Chen Wang, Kequan Zhang, and Haiyan Lu.
\newblock Research and application of a hybrid model based on multi-objective
  optimization for electrical load forecasting.
\newblock {\em Applied Energy}, 180(C):213--233, 2016.

\bibitem{SVR1}
Yongbao Chen, Peng Xu, Yiyi Chu, Weilin Li, Yuntao Wu, Lizhou Ni, Yi~Bao, and
  Kun Wang.
\newblock Short-term electrical load forecasting using the support vector
  regression (svr) model to calculate the demand response baseline for office
  buildings.
\newblock {\em Applied Energy}, 195:659--670, 2017.

\bibitem{HB1}
Nian Liu, Qingfeng Tang, Jianhua Zhang, Wei Fan, and Jie Liu.
\newblock A hybrid forecasting model with parameter optimization for short-term
  load forecasting of micro-grids.
\newblock {\em Applied Energy}, 129:336--345, 2014.

\bibitem{Intro6}
K~Greff, R.~K. Srivastava, J~Koutnik, B.~R. Steunebrink, and J~Schmidhuber.
\newblock Lstm: A search space odyssey.
\newblock {\em IEEE Transactions on Neural Networks and Learning Systems},
  28(10):2222--2232, 2016.

\bibitem{RNN2}
Jiajun Cheng, Pei Li, Zhaoyun Ding, Sheng Zhang, and Hui Wang.
\newblock Sentiment classification of chinese microblogging texts with global
  rnn.
\newblock In {\em IEEE International Conference on Data Science in Cyberspace},
  pages 653--657, 2017.

\bibitem{RNN3}
Kyunghyun Cho, Bart Van~Merrienboer, Caglar Gulcehre, Dzmitry Bahdanau, Fethi
  Bougares, Holger Schwenk, and Yoshua Bengio.
\newblock Learning phrase representations using rnn encoder-decoder for
  statistical machine translation.
\newblock {\em Computer Science}, 2014.

\bibitem{RNN41}
Heng Shi, Minghao Xu, and Ran Li.
\newblock Deep learning for household load forecasting – a novel pooling deep
  rnn.
\newblock {\em IEEE Transactions on Smart Grid}, PP(99):1--1, 2017.

\bibitem{LSTM1}
Alex Graves.
\newblock Long short-term memory.
\newblock {\em Neural Computation}, 9(8):1735--1780, 1997.

\bibitem{LSTM2}
Zheng Zhao, Weihai Chen, Xingming Wu, Peter C.~Y. Chen, and Jingmeng Liu.
\newblock Lstm network: a deep learning approach for short-term traffic
  forecast.
\newblock {\em Iet Intelligent Transport Systems}, 11(2):68--75, 2017.

\bibitem{LSTM3}
Weicong Kong, Zhao~Yang Dong, Youwei Jia, David~J. Hill, Yan Xu, and Yuan
  Zhang.
\newblock Short-term residential load forecasting based on lstm recurrent
  neural network.
\newblock {\em IEEE Transactions on Smart Grid}, PP(99):1--1, 2017.

\bibitem{EMD1}
Norden~E. Huang, Zheng Shen, Steven~R. Long, Manli~C. Wu, Hsing~H. Shih, Quanan
  Zheng, Nai~Chyuan Yen, Chao~Tung Chi, and Henry~H. Liu.
\newblock The empirical mode decomposition and the hilbert spectrum for
  nonlinear and non-stationary time series analysis.
\newblock {\em Proceedings Mathematical Physical and Engineering Sciences},
  454(1971):903--995, 1998.

\bibitem{FT}
K.~W. Cattermole.
\newblock The fourier transform and its applications.
\newblock {\em Electronics and Power}, 11(10):357, 2009.

\bibitem{WT}
I.~Daubechies.
\newblock The wavelet transform, time-frequency localization and signal
  analysis.
\newblock {\em Journal of Renewable and Sustainable Energy}, 36(5):961--1005,
  2015.

\bibitem{RNN}
A.~Rahman, V.~Srikumar, and A.~D. Smith.
\newblock Predicting electricity consumption for commercial and residential
  buildings using deep recurrent neural networks.
\newblock {\em Applied Energy}, 212:372--385, 2018.

\bibitem{PP}
Gakuto Kurata, Bing Xiang, Bowen Zhou, and Mo~Yu.
\newblock Leveraging sentence-level information with encoder lstm for semantic
  slot filling.
\newblock pages 2077--2083, 2016.

\bibitem{GRU}
Mirco Ravanelli, Philemon Brakel, Maurizio Omologo, and Yoshua Bengio.
\newblock Light gated recurrent units for speech recognition.
\newblock {\em IEEE Transactions on Emerging Topics in Computational
  Intelligence}, 2(2):92--102, 2018.

\bibitem{GL}
Nal Kalchbrenner, Ivo Danihelka, and Alex Graves.
\newblock Grid long short-term memory.
\newblock {\em Computer Science}, 2015.

\bibitem{Adam}
Diederik Kingma and Jimmy Ba.
\newblock Adam: A method for stochastic optimization.
\newblock {\em Computer Science}, 2014.

\bibitem{PSO}
Z.~A. Bashir and M.~E. El-Hawary.
\newblock Applying wavelets to short-term load forecasting using pso-based
  neural networks.
\newblock {\em IEEE Transactions on Power Systems}, 24(1):20--27, 2009.

\bibitem{RNN4}
S.~Anbazhagan and N.~Kumarappan.
\newblock Day-ahead deregulated electricity market price forecasting using
  recurrent neural network.
\newblock {\em IEEE Systems Journal}, 7(4):866--872, 2013.

\bibitem{GRU1}
Kuan Lu, Yan Zhao, Xin Wang, Yan Cheng, Xiangkun Pang, Wenxue Sun, Zhe Jiang,
  Yong Zhang, Nan Xu, and Xin Zhao.
\newblock Short-term electricity load forecasting method based on multilayered
  self-normalizing gru network.
\newblock In {\em IEEE Conference on Energy Internet and Energy System
  Integration}, pages 1--5, 2018.

\end{thebibliography}
	
\end{document}